\documentclass[aps,pre,twocolumn]{revtex4-2}
\usepackage{amsmath}
\usepackage{amssymb}

\newcommand \grad  {\mathbf{grad} \, }   
\newcommand \ddiv  {\mathrm{div} \, }   
\newcommand \curl  { \mathbf{curl} \;}   

\begin{document}

\title{Electric field based Poisson-Boltzmann:  Treating mobile charge as polarization}

\author{Michiel Sprik}
\email{ms284@cam.ac.uk}
\affiliation{Department of Chemistry, University of Cambridge,
  Lensfield Road, Cambridge CB2 1EW, United Kingdom}


\begin{abstract}
 Mobile charge in an electrolytic solution can in principle be represented as the divergence of ionic polarization. After adding explicit solvent polarization a finite volume of electrolyte can then be treated as a composite non-uniform dielectric body.  Writing the electrostatic interactions as an integral over electric field energy density we show that the Poisson-Boltzmann functional in this formulation  is convex and can be used to derive the equilibrium equations for electric potential and ion concentration by a variational procedure  developed by Ericksen for dielectric continua  (Arch. Rational Mech. Anal. 2007, {\bf 183}, 299-313).  The Maxwell field equations are enforced by extending the set of variational parameters by a vector potential representing the dielectric displacement which is fully transverse in a dielectric system without embedded external charge. The electric field energy density in this representation is a function of the vector potential and the sum of ionic and solvent polarization making the mutual screening explicit.  
 
\end{abstract}

\maketitle

\section{Introduction}
Poisson-Boltzmann theory continues to be the basis for the understanding of the properties of non-uniform electrolytes.\cite{Hansen2013,Hansen2000,Bazant2007,Haertel2017}.  The main reason for the popularity of   the Poisson-Boltzmann (PB) model is that it captures the essence of the competition between finite temperature entropy and electrostatic interactions determining the structure and energetics of electric double layers. However, PB theory is strictly only valid for dilute solutions. It is missing the short range correlations due to steric interactions between finite size ions as well as long range electrostatic correlations. This has  prompted a huge effort to lift some of these limitations by extending and modifying the original functional\cite{Bazant2007,Podgornik2009,Andelman2019,Kornyshev2020}. The shortcomings of PB theory are also  the motivation behind the subsequent  development of  compact Landau-Ginzburg type functionals for ionic liquids\cite{Bazant2011,Bazant2020}. Of course these issues have also been investigated by methods at the more fundamental end of theory such as integral equation based statistical mechanics\cite{Blum1978,Patey1980,Blum1990,Evans1994,Attard1996}, density functional theory(DFT) \cite{Evans1980,Haertel2017,Haertel2016} and statistical field theory\cite{Netz2001,Podgornik2013,Orland2018}.  Activity in this field has recently received a new impulse\cite{Andelman2019,Kjellander2018,Rotenberg2018,Yochelis2018} by the discoveries made by surface force measurements showing that long range correlations persist  exceeding the PB (Debye) screening length in high concentration electrolytes and ionic liquids\cite{Perkin2016}. 

 None of these profound problems are addressed in the present contribution which focuses instead on a more technical  aspect of classical PB theory namely its status as a variational method.  Clearly the original PB functional is variational, being an elementary example of a density functional\cite{Evans1979}. However this is of limited help in practical calculations  because of the long range character of the electrostatic interaction between charge densities.  The electrostatic energy can however also be expressed as an integral over the square of the Maxwell electric field (with proper attention to  surface terms). The Maxwell field in  turn can be written as the gradient of the same electrostatic potential determining the ion concentration.  An option therefore is  to convert the PB functional in a (semi) local functional of electrostatic potential and ion densities. 
 The Landau-Ginzburg type functionals of Refs.~\citenum{Bazant2011,Bazant2020}  go one step further and express the energy exclusively in terms of the electrostatic potential and its gradients. Treating the electrostatic potential as an auxiliary variational degree of freedom one would hope that the Euler Lagrange equation for the potential is equivalent to the Poisson equation.  Somewhat mysteriously this is not the case (for a review of this issue see Ref.~\citenum{Maggs2012}). The proper Poisson Equation can, however, be generated by a different functional obtained by a Legendre transform. Unfortunately this changes the sign of the field energy from positive to negative. As a result  the stationary solution, while correct,  is no longer a minimum but a saddle point. 

The question of how to restore stability to a field based formulation of the PB functional is  highly relevant for computation relying on iterative methods.  An ingenious solution proposed by Anthony Maggs is to impose the Maxwell equations by means of constraints implemented by the method of undetermined Lagrange multipliers\cite{Maggs2012,Maggs2004}.  The same conflict of the sign of the electrostatic  energy is also encountered in field based variational methods for pure dielectric continua\cite{Hansen2001} and could be resolved using a similar constraint scheme\cite{Maggs2006}.  This method proved particularly suitable for implementation in a molecular dynamics framework\cite{Dunweg2004,Holm2006} and has been applied with success in numerical investigations in colloid science, electrochemistry  and biophysics\cite{Cruz2012,Holm2014}. 

Variational electrostatics is pursued in many diverse disciplines sometimes with little overlap. A development parallel to the activity in the physical chemistry of solutions took place in the field of continuum mechanics of solids\cite{Ericksen2007,Liu2013}. Of particular interest is the approach of J.~L.~Ericksen who reconsidered the use of an extended variational scheme with the dielectric displacement as basis rather than the electric field\cite{Ericksen2007}. The divergence of dielectric displacement vanishes in a dielectric continuum without embedded external charge and can therefore be represented as the curl of a vector potential, just as the magnetic induction. Using a convenient form of the energy functional Ericksen found that under conditions of stationarity for variation of this vector potential the curl of the electric field is zero, as required by the conjugate Maxwell equation\cite{Ericksen2007}. Moreover, being  an expression of the original  dielectric  functional in different variables, the energy functional remains convex.

Ericksen developed his method as part of the continuing effort of merging electrostatics and elasticity theory (For recent reviews see for example Refs.~\citenum{Suo2008} and \citenum{Ogden2017}). A source of inspiration for some of the recent developments in this challenging field is a text book by Kovetz on Maxwell-Lorentz continuum theory of electromagnetism\cite{Kovetz2000}. Maxwell-Lorentz (ML) continuum theory makes no assumptions about the microscopic atomistic nature of polarization. Instead polarization is introduced as the electric component of a vector potential for the internal (material) charge current (the other  component is the magnetization).  This charge-current potential  is designed to satisfy charge conservation by construction and vanishes outside the material body (see also Landau and Lifshitz\cite{Landau1984}). The third defining principle  is the Lorentz equation ($\mathbf{d} = \epsilon_0 \mathbf{e} + \mathbf{p}$) relating polarization $\mathbf{p}$ to the Maxwell field $\mathbf{e}$ and dielectric displacement $\mathbf{d}$. The Lorentz relation  is a fundamental equation and must be considered as an additional field equation  complementing the Maxwell equations for electric and magnetic fields.  The Lorentz relation has the status of a postulate in ML continuum theory, a point forcefully and convincingly made by Kovetz\cite{Kovetz2000}.
 
Tying polarization to internal charge without further atomistic specifications has implications for the interpretation of polarization. This concerns in particular the relation between internal and bound charge. Internal charge in the ML continuum view is bound in the sense that it is confined to the material body but is free to migrate within its boundaries which are assumed to be well defined. Mobile charge in conductors can therefore also be regarded as internal charge. The distinction between charge bound in polar molecules and free charge is made at the level of constitutive equations. This interpretation of internal charge is consistent with the picture presented in  Landau and Lifshitz\cite{Landau1984} and is our justification for treating the ion charge density in electrolytes as the divergence of an ionic polarization. This then allows us to apply variational methods for the electrostatics of dielectric continua to electrolytic solutions.  If the volume of  electrolyte is finite and its boundaries fixed the system should be stable even if charge redistribution is more drastic compared to a conventional dielectric.

  Summarizing, the main purpose of this paper is to show that variational methods for determination of equilibrium polarization in continuum model systems are not limited to dielectric material but can also be used for (confined) conducting ionic systems or to a mixture of the two (electrolytic solutions).  This requires a generalization of the concept of polarization which is already inherent in Maxwell-Lorentz continuum theory. In this context we can also point out that abandoning the narrow definition of polarization as a dipole density is also natural in solid state physics. Partitioning a ionic solid in dipolar units is artificial in finite pieces of crystal and impossible in periodic extended systems\cite{Martin1974,Vanderbilt1993,Resta1994,Resta2007,Resta2010}.  The method is tested out on the Poisson-Boltzmann functional.  Sections \ref{sec:FPB} and \ref{sec:var} are a detailed presentation of the method and a verification that this approach indeed leads to the familiar PB equations for concentration and electrostatic potential. As an illustration of ionic polarization the method is applied in section \ref{sec:polplane} to a planar film of PB electrolyte polarized by a normal external electric field. We conclude in section \ref{sec:disc} with a discussion about the potential of this approach in the theory of electrolytes.  

\section{Free energy functional} \label{sec:FPB}
\subsection{Poisson-Boltzmann theory} \label{sec:Fion}
The Poisson-Boltzmann (PB) free energy consists of two terms\cite{Hansen2013,Hansen2000,Haertel2017}
\begin{equation}
\mathcal{F}_{\textrm{PB}} =  \mathcal{F}_{id}  + \mathcal{F}_{ex}
\label{eqn:FPB}
\end{equation}
$\mathcal{F}_{id}$ is the entropic free energy of non-interacting ions and is referred to as the ideal free energy.
$ \mathcal{F}_{ex} $ is the excess free energy. The primitive variables in  conventional PB theory are the number densities $\nu_{\nu}$ of point-like ion species distinguished by the index $\nu$. For the simple binary system considered here these are the number densities of positive ($\nu=+)$ and negative ($\nu=-$) ions.  The corresponding charge density is written as
 \begin{equation}
\rho(\mathbf{r}) = \sum_{\nu} q_{\nu}  n_{\nu}(\mathbf{r}) 
\label{eqn:rhon}
\end{equation}
$q_{+} = q>0$ is the charge of the cation and $q_{-} = - q $ the charge of the anion.  The ionic liquid is overall neutral
\begin{equation}
   \int_{\Omega} \rho(\mathbf{r}) dv = 0 
\label{eqn:neutral}
\end{equation}
where $\Omega$ is the volume of the ``body'' of ionic solution. $dv = d r^3$ is an elementary volume element. 
 
 The ideal free energy $ \mathcal{F}_{id} $  is the sum of the (gas phase) translational free energy of the mixture of ionic species
\begin{equation}
\mathcal{F}_{id} =  \int_{\Omega} \sum_{\nu} f\left(n_{\nu}(\mathbf{r})\right)  dv
\equiv \mathcal{F}_{m} [n_{\nu}]
\label{eqn:Fid}
\end{equation}
with the local free energy density $f(n)$ given by 
\begin{equation}
 f(n_{\nu}) = k_{\textrm{B}} T  \left( n_{\nu} \ln \left(n_{\nu} \Lambda^3 \right) - n_{\nu} \right)
 \label{eqn:fn}
\end{equation} 
where $ k_{\textrm{B}}$ is Boltzmann's constant and $T$ the temperature. $\Lambda$ is the thermal wavelength (the ions have the same mass).  The number densities (``absolute'' concentrations) are related to  $f(n_{\nu})$ by the ideal chemical potential 
\begin{equation}
\mu_{\nu}^c = \frac{ \partial f(n_{\nu})}{\partial n_{\nu}} 
 =  k_{\textrm{B}} T \ln \left( n_{\nu} \Lambda^3 \right)
\label{eqn:munu}
\end{equation}
In applications concentrations are usually referred to a convenient  standard concentration.

The excess free energy of PB theory is the mean field electrostatic energy $\mathcal{E}_{\textrm{C}} $ 
\begin{equation}
 \mathcal{E}_{\textrm{C}} = \frac{1}{8 \pi \epsilon} \int_{\Omega}
 \frac{\rho(\mathbf{r}) \rho(\mathbf{r^\prime})}{ | \mathbf{r} - \mathbf{r^\prime}|}
  dv dv^\prime \equiv \mathcal{E}_{\rho} [\rho] 
\label{eqn:Erho}
\end{equation}
expressed a Coulomb integral of the charge density. $\epsilon$ is the permittivity of the solvent approximated by a polarizable continuum.  Adding the functionals defined in Eq.~\ref{eqn:Fid}  and \ref{eqn:Erho}  gives the well-known Poisson-Boltzmann density functional $\mathcal{F}_{i} [n_{\nu}]$ used in the literature 
\begin{equation}
\mathcal{F}_{\textrm{PB}} =  \mathcal{F}_{m} [n_{\nu}] + \mathcal{E}_{\rho} [\rho]  \equiv
\mathcal{F}_{i} [n_{\nu}]  
 \label{eqn:Fi}
\end{equation}

\subsection{Mobile charge as polarization} \label{sec:Fdiel}
The PB functional Eq.~\ref{eqn:Fi} will be modified and extended for the particular variational treatment proposed here. First we change variables to total (number) density $n$ 
\begin{equation}
   n =   n_{+} + n_{-}  
   \label{eqn:sumn} \\
\end{equation}
 and the charge density represented in terms of an ionic polarization $\mathbf{p}_i$
\begin{equation}
  \ddiv \mathbf{p}_i  =  - \rho
 \label{eqn:pion}
\end{equation}
In this new set of primitive variables the ionic densities are expressed as 
\begin{equation}
 n_{\nu}\left(n,\mathbf{p}_i\right)
  =   \frac{1}{2} \left( n - \frac{1}{q_{\nu}} \ddiv \mathbf{p}_i \right)
\label{eqn:nunp}
\end{equation}
Eq.~\ref{eqn:pion} is a formal definition of the polarization associated with mobile charge. The physical interpretation  of  $\mathbf{p}_i$  will concern us later when applying the transformed PB functional to the example of a polarized planar layer of electrolyte. 

Treating mobile charge as internal charge has immediate implications for the dielectric displacement field $\mathbf{d}$ inside the electrolyte. Without embedded external charge (as distinct from free charge) $\mathbf{d}$ is transverse .
\begin{equation}
\ddiv \mathbf{d} = 0
\label{eqn:divd}
\end{equation}
This property, characteristic of dielectrics, will be exploited to design a field based variant of PB theory. The Maxwell electric field  $\mathbf{e}$ of course still satisfies the general Maxwell equation for (static) electric fields 
\begin{equation}
\curl \mathbf{e} = 0
\label{eqn:curle}
\end{equation}
and can, as usual, be represented as (minus) the gradient of an electrostatic potential. Dielectric displacement and Maxwell field are related to the ionic polarization by the  Lorentz relation
\begin{equation}
\mathbf{d} = \epsilon_0 \mathbf{e} + \mathbf{p}_i
\label{eqn:Lorentz}
\end{equation}
Note that $\curl \mathbf{p}_i =  \curl \mathbf{d}$ may be finite. As in dielectric theory we expect the transverse component of $\mathbf{p}_i$ to be determined by  constitutive (material) relations. 

 Substituting Eq.~\ref{eqn:nunp} in Eq.~\ref{eqn:Fid} we obtain our alternative  expression for $\mathcal{F}_{id}$ 
\begin{equation}
 \mathcal{F}_{id} = \int_{\Omega} \sum_{\nu} f\left(n_{\nu}(n, \mathbf{p}_i) \right) dv
 \equiv \mathcal{F}_{pi} [n, \mathbf{p}_i]
\label{eqn:Fipol}
\end{equation}
which will take over the role of ideal free energy.  

Having written the ideal free energy  in terms of total density $n$ and polarization $\mathbf{p}_i$ we next  cast the electrostatic interaction energy $\mathcal{E}_{\textrm{C}}$ in a similar form. Following  Ericksen  we do this by switching to a field representation\cite{Ericksen2007}. 
\begin{equation}
\mathcal{E}_{\textrm{C}} = \int_V \frac{\epsilon_0\mathbf{e}^2}{2} dv \equiv \mathcal{E}_F
 \label{eqn:EML}
\end{equation}
 where $\mathbf{e}$ is the Maxwell electric field. $\mathcal{E}_F$ is the total field energy of Maxwell-Lorentz continuum theory\cite{Kovetz2000}. This is why the dielectric constant in Eq.~\ref{eqn:EML} is the permittivity $\epsilon_0$ of vacuum. A further comment concerns the system geometry. The ionic solution is assumed to be a finite volume of material placed in a container of total volume $V$ much larger than the body volume $\Omega$.  While the induced charge density $\rho$ is confined to the body, the electric field generated by the excess charge spills out into the surrounding region of the container (assumed to be vacuum).   The integral over $\Omega$ in Eq.~\ref{eqn:Erho} had therefore to be extended to an integral over $V$ in Eq.~\ref{eqn:EML}. This raises the infamous issue of surface terms in electrostatics to which we return in section \ref{sec:efield}.
 
 The field energy $\mathcal{E}_F$ represents  the sum total of all electrostatic interactions, ion-ion,  ion-solvent, solvent-solvent and the interaction with the external polarizing device. Even the self energy of the external field  is  included (which will be taken out in section \ref{sec:efield}).   However,  $\mathcal{E}_F$ does not account for the polarization energy of the solvent which will have to be added as a separate contribution $ \mathcal{F}_{pd}$ to the excess free energy. 
 \begin{equation}
 \mathcal{F}_{pd} = \int_{\Omega} \left(\frac{\mathbf{p}_d^2}{2 \chi} \right) dv
 \label{eqn:Fdpol}
\end{equation}
where $\mathbf{p}_d$ is the polarization of the solvent and  $\chi$ the solvent susceptibility.  $\mathcal{F}_{pd}$ of Eq.~\ref{eqn:Fdpol} is a constitutive free energy quantifying the cost in energy  of polarizing the solvent other than the electrostatic energy which is part of the field energy Eq.~\ref{eqn:EML}. For a simple polar fluid consisting of  rigid dipoles centered on atoms (Stockmayer fluid)  $\mathcal{F}_{pd}$  is largely entropic.  This suggests that  $\mathcal{F}_{pi}$ of Eq.~\ref{eqn:Fipol} can be interpreted as a similar constitutive  energy for polarization of  the ionic fluid. Added together these energies define a constitutive energy 
\begin{equation}
\mathcal{F}_S  = \mathcal{F}_{pi}\lbrack n, \mathbf{p}_i \rbrack +\mathcal{F}_{pd}\lbrack \mathbf{p}_d \rbrack
\label{eqn:Fs}
\end{equation}
which in solid mechanics is often referred to as the stored energy. However, where it concerns the electrostatics there is no fundamental difference  between  $\mathbf{p}_i$ and $\mathbf{p}_d$. Both  are polarization and are therefore superimposed in the Lorentz relation. Hence instead of  Eq.~\ref{eqn:Lorentz} we have  
 \begin{equation}
\mathbf{d} = \epsilon_0\mathbf{e} + \mathbf{p}_i + \mathbf{p}_d
\label{eqn:Lorentzid}
\end{equation}
As a consequence the field energy $\mathcal{E}_F$ of Eq.~\ref{eqn:EML} is determined by the sum of ionic and solvent polarization ($\mathcal{E}_{\textrm{C}} =\mathcal{E}_F \lbrack\mathbf{p}_i + \mathbf{p}_d \rbrack$). This is a central feature of our approach and will be demonstrated in detail in section \ref{sec:efield}.   

 Summarizing, the statement we make is that the standard expression Eq.~\ref{eqn:Fi} for the Poisson-Boltzmann functional can be rewritten in ``pseudo'' dielectric  form
\begin{equation}
\mathcal{F}_d [n, \mathbf{p}_i,\mathbf{p}_d]  = \mathcal{F}_S [n, \mathbf{p}_i,\mathbf{p}_d] + 
 \mathcal{E}_{F} [\mathbf{p}_i + \mathbf{p}_d] 
 \label{eqn:Fd}
\end{equation}
where  $\mathcal{F}_S$ is the constitutive energy functional of Eq.~\ref{eqn:Fs}.  The precise  expression for the electrostatic field energy $\mathcal{E}_{F}$ will be given below. Total density $n$ and ionic and solvent polarization $\mathbf{p}_i,\mathbf{p}_d$ are treated as independent variational degrees of freedom.   The dielectric displacement $\mathbf{d}$ and electric field $\mathbf{e}$ again must satisfy the Maxwell equations   Eq.~\ref{eqn:divd} respectively Eq.~\ref{eqn:curle} but now with Eq.~\ref{eqn:Lorentzid} acting as the third field equation. The functional Eq.~\ref{eqn:Fd} with its extended set of variables describes strictly the same PB electrolyte as Eq.~\ref{eqn:Fi}.  A possible dependence of susceptibility on the solvent density has therefore been ignored. The volume of the solution is assumed to be kept in shape  by a rigid electrically inert wall. Jump conditions for $\mathbf{d}$ and $\mathbf{e}$ of course apply.
 
\subsection{Pseudo dielectric  field energy} \label{sec:efield}
 
 The electrostatic field energy $\mathcal{E}_F$ of Eq.~\ref{eqn:EML} was converted by Ericksen\cite{Ericksen2007} into an expression appropriate and convenient for a finite body of dielectric material subject to an external electric  field $\mathbf{e}_0$.  The system responds with a self field $\hat{\mathbf{e}}$ (indicated by the hat). The two add up to the Maxwell field
 \begin{equation}
  \mathbf{e} = \mathbf{e}_0 + \hat{\mathbf{e}}
  \label{eqn:eMax}
\end{equation}
which is the field determining the electrostatic energy $\mathcal{E}_F$ of  Eq.~\ref{eqn:EML}.   $\mathcal{E}_F$ includes the self energy of the vacuum field $\mathbf{e}_0$. 
 Separating this energy out we can write
\begin{equation} 
\mathcal{E}_F = \int_V \frac{\epsilon_0\mathbf{e}_0^2}{2} dv + \mathcal{U}_F
\label{eqn:UML}
\end{equation}
$\mathcal{U}_F$ is the system energy we are interested in.  It should vanish when the electrically active system is removed. Similar to Eq.~\ref{eqn:EML} the internal energy $\mathcal{U}_F$  can be written  as an integral over a field energy density
\begin{equation}
\mathcal{U}_F = \int_{V} e_{\textrm{E}}(\mathbf{p}) dv \equiv \mathcal{E}_{\textrm{E}}[\mathbf{p}]
 \label{eqn:Ediel}
\end{equation}
with $e_{\textrm{E}}$ given by
\begin{equation}
  e_{\textrm{E}}(\mathbf{p}) = -\mathbf{p} \cdot \mathbf{e}_0 + \frac{\epsilon_0\hat{\mathbf{e}}^2 }{2}
  \label{eqn:Ekel} 
\end{equation}
Note that the dielectric constant is still the permittivity of vacuum as in Eq.~\ref{eqn:EML}.

In Ref.~\citenum{Ericksen2007} Ericksen derived Eq.~\ref{eqn:Ekel} for a  piece of dielectric solid with a possibly non-uniform polarization $\mathbf{p}(\mathbf{r})$.  For the application to PB theory $\mathbf{p}$ will be generalized to the composite electrolyte polarization appearing  in the Lorentz relation Eq.~\ref{eqn:Lorentzid}.  Therefore, $\mathbf{p}$ in Eq.~\ref{eqn:Ekel} is set equal to
 \begin{equation}
\mathbf{p} = \mathbf{p}_i + \mathbf{p}_d
\label{eqn:pel}
\end{equation}
In molecular physical chemistry  Eq.~\ref{eqn:Ekel} is the ubiquitous expression for the energy of  a dipole distribution in an applied field. The reason that the generalization Eq.~\ref{eqn:pel} is allowed is that  ML continuum theory does not discriminate between different types of polarization\cite{Kovetz2000}. This distinction is made by constitutive relations, Eq.~\ref{eqn:Fipol} and \ref{eqn:Fdpol} for our system.  The ultimate justification for this dielectric view of an electrolyte is that we are able to reproduce the established results for PB theory. However,   it is instructive to briefly repeat the derivation of Ref.~\citenum{Ericksen2007} to emphasize that Eq.~\ref{eqn:Ekel} is also valid for a conductor, provided the material system is finite.

 Expanding the square of $\mathbf{e}  = \mathbf{e}_0 + \hat{\mathbf{e}}$ in the field energy density
Eq.~\ref{eqn:EML}
\begin{equation}
 \frac{\epsilon_0}{2} \mathbf{e}^2 = \frac{\epsilon_0}{2} \mathbf{e}_0^2
  + \epsilon_0 \mathbf{e_0} \cdot \hat{\mathbf{e}} + \frac{\epsilon_0}{2} \hat{\mathbf{e}}^2
\end{equation}
and comparing to Eq.~\ref{eqn:Ekel} we see that the difference is in the cross term  $ \epsilon_0 \mathbf{e_0} \cdot \hat{\mathbf{e}}$. This term is converted using the property Eq.~\ref{eqn:divd}  of the displacement field of a pure dielectric. Instead of an external charge distribution the system is polarized by  the applied vacuum field  $\mathbf{e}_0$  with again $\ddiv \mathbf{e}_0 = 0$.  The self displacement field $\hat{\mathbf{d}}$ associated with the self field 
\begin{equation}
     \hat{\mathbf{d}} =  \epsilon_0 \hat{\mathbf{e}} + \mathbf{p} 
\label{eqn:hatLorentz} 
\end{equation}
is therefore also transverse
\begin{equation}
 \ddiv \hat{\mathbf{d}} = 0
 \label{eqn:Maxd}
\end{equation}
Equation Eq.~\ref{eqn:Maxd} together with the Maxwell equation for the self field 
 \begin{equation}
 \curl \hat{\mathbf{e}} = 0
 \label{eqn:Maxe}
\end{equation}
is therefore equivalent to  Eqs.~\ref{eqn:divd} and \ref{eqn:curle}. 

Substituting Eq.~\ref{eqn:hatLorentz} in the cross term integral
\begin{equation}
\int_V \epsilon_0 \mathbf{e_0} \cdot \hat{\mathbf{e}} \, dv =
\int_V \mathbf{e_0} \cdot \hat{\mathbf{d}} \, dv
- \int_\Omega \mathbf{e_0} \cdot \mathbf{p}  \, dv
\label{eqn:inte0e}
\end{equation}
and applying the divergence theorem we obtain for the first term with $\mathbf{e}_0 = - \grad \phi_0$ 
\begin{equation}
 \int_V \mathbf{e_0} \cdot \hat{\mathbf{d}} \, dv  =
  - \int_{\partial V}  \phi_0 \, \hat{\mathbf{d}} \cdot  \mathbf{n}_{\partial V} \, ds +
 \int_V  \phi_0 \ddiv \hat{\mathbf{d}} \, dv
 \label{eqn:inte0d} 
\end{equation}
where $\partial V$ is the boundary of $V$ with normal $\mathbf{n}_{\partial V}$. 
Because the system carries no net charge (Eq.~\ref{eqn:neutral}) the self  displacement can be assumed to decay to zero sufficiently fast so it can be neglected at the boundary of the container. The surface integral in Eq.~\ref{eqn:inte0d} vanishes. So does the volume integral because of Eq.~\ref{eqn:Maxd}.
What is left of Eq.~\ref{eqn:inte0e} is  only the second term which becomes the coupling between the external field $\mathbf{e_0}$ and the polarization $\mathbf{p}$ in Eq.~\ref{eqn:Ekel}.  Note that we have arrived at this result bypassing multipole expansions, neither locally in the form of a dipole density or globally. The minimal condition that the electrically responsive  material occupies a finite volume and is neutral was sufficient. This brings us finally to the question of stability. The original full field energy  $\mathcal{E}_F$ of Eq.~\ref{eqn:EML} is manifestly convex.   Because the field energy $\mathcal{E}_{\textrm{E}}$  of Eq.~\ref{eqn:Ediel} differs from  $\mathcal{E}_F$, for  the body in container geometry,  by the fixed energy of the applied field, $\mathcal{E}_{\textrm{E}}$  is also  convex.

\subsection{Vector potential for displacement} \label{sec:vecpot}
  The charge redistribution of an electrolyte in response to applied electric fields $\mathbf{e}_0$ is non-uniform. However, if the system is finite, the currents  should eventually relax to zero with the system reaching an equilibrium state. For a good conductor this might mean that essentially all induced charge will accumulate at the surface.  The equilibrium state is found by minimizing the free energy Eq.~\ref{eqn:Fd} in $n$ and $\mathbf{p}_i, \mathbf{p}_d$ under constraints of the two Maxwell equations Eq.~\ref{eqn:Maxd} and \ref{eqn:Maxe} in combination with the Lorentz relation Eq.~\ref{eqn:hatLorentz}.  As argued earlier, from a technical point of view, there should in principle be no difference between an inhomogeneous  dielectric and an electrolyte and we can proceed using the variational method of Ref.~\citenum{Ericksen2007} for dielectric systems. Ericksen imposes Eq.~\ref{eqn:Maxd} by writing the self displacement $\hat{\mathbf{d}}$  as the curl of a vector potential $\hat{\mathbf{a}}$ 
\begin{equation}
  \hat{\mathbf{d}} = \curl \hat{\mathbf{a}}
  \label{eqn:dcurla}
\end{equation}
The vector field $\hat{\mathbf{a}}$ is treated as a further independent  electric variational parameter in addition to polarization. 

The electrostatic energy density Eq.~\ref{eqn:Ekel} becomes a two-variable function $\tilde{e}_{\textrm{E}}(\mathbf{p}, \hat{\mathbf{a}})$ and is found by substituting Eq.~\ref{eqn:dcurla}  using Eq.~\ref{eqn:hatLorentz}.
\begin{equation}
 \tilde{e}_{\textrm{E}}(\mathbf{p}, \hat{\mathbf{a}}) = -\mathbf{p} \cdot \mathbf{e}_0 + 
 \frac{\left(\curl \hat{\mathbf{a}} - \mathbf{p}\right)^2 }{2\epsilon_0}
  \label{eqn:Ekela}
\end{equation}
with the corresponding extended electrostatic energy functional
\begin{equation}
\tilde{\mathcal{E}}_{\textrm{E}}[\mathbf{p}, \hat{\mathbf{a}}]
 = \int_V \tilde{e}_{\textrm{E}}  (\mathbf{p}, \hat{\mathbf{a}})\, dv
 \label{eqn:Ekintela}
\end{equation}
The polarization $\mathbf{p}$ in Eqs.~\ref{eqn:Ekela} and \ref{eqn:Ekela} is understood to be the composite electrolyte polarization $\mathbf{p}_i + \mathbf{p}_d$ of Eq.~\ref{eqn:pel}.  This is how $\mathbf{p}$ must be read whenever it appears in the following.
 
The three variable energy functional Eq.~\ref{eqn:Fd} is extended to a four variable functional
 \begin{equation}
  \tilde{\mathcal{F}}_{d}[n, \mathbf{p}_i, \mathbf{p}_d, \hat{\mathbf{a}}]  =  
 \mathcal{F}_S[n, \mathbf{p}_i,\mathbf{p}_d] 
  + \tilde{\mathcal{E}}_{\textrm{E}}[\mathbf{p}, \hat{\mathbf{a}}]
  \label{eqn:Fepa}
\end{equation}
 $\mathbf{p}_i, \mathbf{p}_d$ and $\hat{\mathbf{a}}$ are independent variational degrees of freedom. The constitutive function $\mathcal{F}_S$ is therefore independent of $\hat{\mathbf{a}}$  and is still given by Eqs.~\ref{eqn:Fipol} and \ref{eqn:Fdpol}(so no tilde).  The central idea of the Ericksen procedure is that the Euler-Lagrange equation for the vector potential  $\hat{\mathbf{a}}$  will generate the Maxwell equation Eq.~\ref{eqn:Maxe} for the self field, which will be verified in the next section.

\section{Variational procedure} \label{sec:var}

\subsection{Varying the vector potential} \label{sec:varvec}
 Changing $\hat{\mathbf{a}}$ to $\hat{\mathbf{a}} + \delta \hat{\mathbf{a}}$ keeping $n$ and $\mathbf{p}$ fixed yields a first order change in the field energy
\begin{equation}
\delta \tilde{\mathcal{E}}_{\textrm{E}} = 
 \frac{1}{\epsilon_0} \int_V \left( \curl \hat{\mathbf{a}} - \mathbf{p}\right)  \cdot \curl \delta \hat{\mathbf{a}} \, dv
\end{equation}
where we have used that $ \delta \left(\curl \hat{\mathbf{a}} \right) = \curl \delta \hat{\mathbf{a}}$.  Rewriting the integrand  using the vector identity
\begin{equation} 
\ddiv\left( \mathbf{u} \times \mathbf{v} \right) = \mathbf{v} \cdot \left( \curl \mathbf{u} \right) - \mathbf{u}  \cdot \left( \curl \mathbf{v} \right)
\label{eqn:divuoutv}
\end{equation}
we can apply the divergence theorem and obtain
\begin{align}
\delta \tilde{\mathcal{E}}_{\textrm{E}} & = \frac{1}{\epsilon_0} \int_V  \curl \left( \curl\hat{\mathbf{a}} - \mathbf{p}\right)  \cdot \delta \hat{\mathbf{a}} \, dv  \nonumber \\ &
   + \frac{1}{\epsilon_0} \int_{\partial V}  \mathbf{n}_{\partial V} \times \left( \curl \hat{\mathbf{a}} - \mathbf{p}\right)  \cdot \delta \hat{\mathbf{a}} \, ds
   \label{eqn:dwall}
\end{align}
While $\mathbf{p}$ is strictly zero beyond the periphery of a finite dielectric body  the vector potential $\hat{\mathbf{a}}$, similar to $\hat{\mathbf{e}}$ is not. However,  similar to the surface integral in Eq.~\ref{eqn:inte0d} we can expect that  $\hat{\mathbf{a}}$ decays to zero with increasing distance from the dielectric body and can be neglected at the vacuum boundary $\partial V$ leaving only the spatial integral over  $V$.  From Eq.~\ref{eqn:Fepa} we know that $\delta  \tilde{\mathcal{F}}_{d} =\delta \tilde{\mathcal{E}}_{\textrm{E}}$ for variation in $\hat{\mathbf{a}}$. Hence, we can apply the usual argument in variational theory and require that the  integral Eq.~\ref{eqn:dwall}  must vanish for arbitrary $\delta \hat{\mathbf{a}}$.  This is only possible if
\begin{equation}
\curl \left(\curl\hat{\mathbf{a}} - \mathbf{p}\right) = 0
\label{eqn:curl2a}
\end{equation} 
validating  the identification 
\begin{equation}
 \hat{\mathbf{e}} = \left( \curl \hat{\mathbf{a}} - \mathbf{p} \right)/\epsilon_0
\label{eqn:eselfa}
\end{equation}
 with $\hat{\mathbf{e}} $ satisfying Eq.~\ref{eqn:Maxe}. 
 
 Similar to the vector potential for magnetic induction,  Eq.~\ref{eqn:dcurla} leaves  us with the freedom of choosing a convenient gauge for $\hat{\mathbf{a}}$.   The double curl in Eq.~\ref{eqn:curl2a} suggests an even closer parallel to magnetic induction\cite{Ericksen2007}.  Using another relation from vector analysis this equation can also be written as
 \begin{equation}
 \nabla \ddiv \hat{\mathbf{a}} - \Delta \hat{\mathbf{a}} = \curl \mathbf{p}
 \end{equation}
 where $\Delta$ is the Laplacian differential operator.  We have therefore the option of setting $\ddiv \hat{\mathbf{a}} = 0$ making the vector potential fully transverse. Note that in this gauge   $\Delta \hat{\mathbf{a}} =0$ for systems with only longitudinal  polarization.  Transverse polarization acts as a source for  $\hat{\mathbf{a}}$.

\subsection{Varying polarization}
Variation in the ionic polarization $\mathbf{p}_i \rightarrow \mathbf{p}_i + \delta \mathbf{p}_i$ is carried at fixed vector potential $\hat{\mathbf{a}}$ and solvent polarization $\mathbf{p}_d$.  Holding $\hat{\mathbf{a}}$ constant  greatly simplifies the expression for the first order change of the electrostatic energy
\begin{equation}
\delta \tilde{\mathcal{E}}_{\textrm{E}}  = \int_{\Omega} \left(  - \mathbf{e}_0 -\frac{\left(\curl \hat{\mathbf{a}} - \mathbf{p} \right)}{\epsilon_0}  \right)\cdot \delta \mathbf{p}_i \, dv
\end{equation}
Integration can be limited to $\Omega$ because the polarization remains confined to the body. With no further differentials to determine we can substitute Eq.~\ref{eqn:eselfa}. This gives together with Eq.~\ref{eqn:eMax} for the variation in the field energy 
\begin{equation}
\delta \tilde{\mathcal{E}}_{\textrm{E}}  = - \int_{\Omega} 
\mathbf{e} \cdot \delta \mathbf{p}_i  \, dv
\label{eqn:dteEi}
\end{equation}
The electric field conjugate to $\mathbf{p}_i$ is the full Maxwell field.

 To this we must add the first order change in the constitutive function $\mathcal{F}_{pi} $ keeping total number density $n$ fixed. Expanding Eq.~\ref{eqn:Fipol} in variations of the separate species densities yields
\begin{equation}
\delta \mathcal{F}_{pi} = \int_{\Omega}  \sum_{\nu} \mu_{\nu}^c \delta n_{\nu} dv
\label{eqn:dFPmu}
\end{equation}
where we have substituted Eq.~\ref{eqn:munu}. The variations in ion densities $n_{\nu}$ at constant overall density $n = n_{+} + n_{-}$ are according to Eq.~\ref{eqn:nunp} 
\begin{equation}
 \delta n_{\nu}  =   - \frac{1}{2q_{\nu}} \ddiv \delta \mathbf{p}_i, \qquad
 \label{eqn:dnunp}
\end{equation}
Inserting we find
\begin{equation}
\delta \mathcal{F}_{pi}  =  - \int_{\Omega} \frac{1}{2q} \left( 
  \mu_{+}^c - \mu_{-}^c \right) \ddiv \delta \mathbf{p}_i \, dv 
\end{equation}
To balance $\delta \mathcal{F}_{pi}$ against  variation in field energy Eq.~\ref{eqn:dteEi} for arbitrary $\delta \mathbf{p}_i$  we apply partial integration. This gives assuming again that the surface term can be made to vanish by  extending it into the vacuum container 
 \begin{equation}
\delta \mathcal{F}_{pi}  =   \int_{\Omega} \frac{1}{2q} \nabla \left( 
  \mu_{+}^c - \mu_{-}^c \right) \cdot \delta \mathbf{p}_i \, dv 
\end{equation}
leading with Eq.~\ref{eqn:dteEi} the Euler-Lagrange equation
\begin{equation}
\frac{1}{2q} \nabla \left( \mu_{+}^c - \mu_{-}^c \right) =  \mathbf{e}
\label{eqn:pbgradmu}
\end{equation}
or in terms of concentrations
\begin{equation}
\frac{k_{\textrm{B}} T}{2 q}  \nabla\ln \left[ \frac{n_+}{n_-} \right] = \mathbf{e}
\label{eqn:qnernst}
\end{equation}
where we have used Eq.~\ref{eqn:munu}.

Varying solvent polarization $\mathbf{p}_d \rightarrow \mathbf{p}_d + \delta \mathbf{p}_d$ at fixed $\hat{\mathbf{a}}, \mathbf{p}_i$  and $n$ proceeds along the same line. Since, for electrostatic considerations,  ionic and solvent polarization are additive (Eq.~\ref{eqn:pel}) the first order difference is the same as Eq.~\ref{eqn:dteEi} with $\delta \mathbf{p}_d$ replacing $\delta \mathbf{p}_i$
\begin{equation}
\delta \tilde{\mathcal{E}}_{\textrm{E}}  = - \int_{\Omega} 
\mathbf{e} \cdot \delta \mathbf{p}_d  \, dv
\label{eqn:dteEd}
\end{equation}
The variation in the solvent polarization  constitutive energy is straight forward
 \begin{equation}
\delta \mathcal{F}_{pd} = \int_{\Omega} 
\frac{\mathbf{p}_d}{\chi} \cdot \delta \mathbf{p}_d  \, dv
\end{equation}
The resulting Euler-Lagrange equation is the susceptibility relation for linear dielectrics
\begin{equation}
  \mathbf{p}_d = \chi \mathbf{e}
  \label{eqn:pchie}
\end{equation}
  Note that $\mathbf{e}$ in Eqs.~\ref{eqn:pbgradmu} and \ref{eqn:pchie} is the same. While  polarizations are partial, there is only a single Maxwell field.

\subsection{Varying total density}
The final step is the variation of  total density $n \rightarrow n + \delta n$ carried out at fixed $\mathbf{p}_i, \mathbf{p}_d$ and $\hat{\mathbf{a}}$.  In addition we must ensure that the total number of particles is conserved. This adds a Lagrange multiplier $\mu$ to the Euler-Lagrange equation for $n$. Omitting an external one-particle potential we therefore have the equilibrium condition
\begin{equation}
\frac{\partial \mathcal{F}_{d}}{\partial n} = \mu  
\end{equation}
 The electrostatic energy density of Eq.~\ref{eqn:Ekel} is not affected by changes in $n$.  Neither is the stored polarization energy $\mathcal{F}_{pd}$ of Eq.~\ref{eqn:Fdpol} (This will change if we admit electrostriction). $\delta \mathcal{F}_{d}$ is therefore entirely determined by the differential of the ionic constitutive energy $\delta \mathcal{F}_{pi}$ of Eq.~\ref{eqn:Fipol}.  Keeping ionic polarization constant the change in species densities is according to Eq.~\ref{eqn:nunp} simply $\delta n_{\nu} = \delta n/2$, the same for both species.  Substituting we find 
\begin{equation}
  \delta \mathcal{F}_{pi} = \int_{\Omega}  \frac{1}{2}\left( \mu_{+}^c  + \mu_{-}^c  \right)
  \delta n \, dv
\end{equation}
yielding the equilibrium equation 
\begin{equation}
   \frac{1}{2}\left( \mu_{+}^c  + \mu_{-}^c  \right) = \mu
\label{eqn:ELn}
\end{equation}
Substituting Eq.~\ref{eqn:munu} this is equivalent to
\begin{equation}
\sqrt { n_{+} n_{-} } = \Lambda^{-3} \exp \left[\frac{\mu}{k_{\textrm{B}}T} \right]
\label{eqn:meanact}
\end{equation}
As expected $\mu$ can be identified with the uniform (absolute) mean activity. 

Since $\mu$ is a constant another implication of Eq.~\ref{eqn:ELn} is that $ \nabla \mu_{+}^c  = - \nabla \mu_{-}^c$.  With Eqs.~\ref{eqn:pbgradmu} and  Eq.~\ref{eqn:munu} this yields
\begin{equation}
 \frac{k_{\textrm{B}}T}{q_{\nu}} \nabla \ln \left[\Lambda^{3} n_{\nu}\right] = \mathbf{e}
\end{equation}
Setting  $ \mathbf{e} =  - \nabla \phi$,  integrating and exponentiating the we find for the equilibrium concentrations
\begin{equation}
   n_{\nu} =\frac{n_0}{2} \exp \left[ - \frac{q_{v} \phi}{k_{\textrm{B}}T} \right]
   \label{eqn:pbn0}
\end{equation}
$n_0$ is the total density of  ions at locations  of zero potential.   Using Eq.~\ref{eqn:meanact} we can write  Eq.~\ref{eqn:pbn0} in the more thermochemical form
 \begin{equation}
   n_{\nu} = \Lambda^{-3} \exp \left[ - \frac{\left( q_{v} \phi - \mu \right)}{k_{\textrm{B}}T} \right]
   \label{eqn:pbmu}
\end{equation}
Eq.~\ref{eqn:pbn0} is the crucial result necessary to verify  consistency between the  conventional formulation of Poisson-Boltzmann theory and the polarization based approach developed here.

The final check is reproducing the PB equation for the electrostatic potential $\phi$.  Starting from the  Lorentz relation Eq.~\ref{eqn:Lorentzid} we insert the constitutive Eq.~\ref{eqn:pchie} for the solvent polarization. The differential equation for the potential is obtained from the divergence of this equation.  This is the usual procedure. However in our  ``pseudo dielectric ``  $\ddiv  \mathbf{d} = 0 $ because we have represented the  free charge by the ionic polarization $\mathbf{p}_i$.   We end up with  effectively the identical equation   
\begin{equation}
 \ddiv \left(\epsilon  \,\mathbf{e} \right) = - \ddiv \mathbf{p}_i
\end{equation}
where we have set $\epsilon_0 + \chi = \epsilon $. The right hand side is according to the definition of the ionic polarization (Eq.~\ref{eqn:pion}) the mobile charge density $\rho$.   Substituting the Boltzmann relation 
Eq.~\ref{eqn:pbn0} in Eq.~\ref{eqn:rhon} for the density yields the familiar PB equation for the potential.
  Then, in the conventional notation for the Laplacian $ \Delta \phi = -\ddiv \mathbf{e}$ we obtain the PB equation for the potential\cite{Hansen2013}
\begin{equation}
 \Delta \phi = \frac{qn_0}{\epsilon}  \sinh  \left[  \frac{q\phi}{k_{\textrm{B}}T} \right]
  \label{eqn:phiPB}
\end{equation}
Perhaps it is appropriate to conclude this section with the warning that the self field $\hat{\mathbf{e}}$ determining the electrostatic energy in Eq.~\ref{eqn:Ekel} cannot be replaced by $-\nabla \phi$ because $\phi$ is the total electric potential including the potential for the applied field $\mathbf{e}_0$. In fact $\mathbf{e}_0$ nowhere explicitly appears  in the equilibrium equations. We will come back to this issue in the application presented in the next section.

\section{Planar polarized film} \label{sec:polplane}
\subsection{Linearized Poisson-Boltzmann} \label{sec:linPB}
The example discussed in every text book is a semi infinite volume of solution in contact with a charged wall\cite{Hansen2013}. This is the original Gouy-Chapman (GC) model of an electric double layer. We will investigate a different system more suitable for illustrating the nature of ionic polarization.  This a planar film of finite width $l$ subject to a normal external electric field. Choosing the $z$ axis as the direction perpendicular to the  layer the boundary planes are at $z = \pm l/2$.  These planes carry no surface charge. The layer of electrolyte is polarized by an external electric field of magnitude $e_0$ pointing along the  positive  $z$ axis ($ e_{0z} =  e_0, e_{0x}= e_{0y} = 0$). 

The response of the electrolyte will be studied solving the linearized PB equation Eq.~\ref{eqn:phiPB}. 
\begin{equation}
\frac{d^2 \phi^*(z)}{ d z^2} = k_{\textrm{D}}^2 \phi^*(z)
\label{eqn:PBlin}
\end{equation}
where $\phi^* (z)= q \phi(z)/(k_{\textrm{B}}T)$ is the dimensionless potential and $k_{\textrm{D}}$ is the inverse Debye screening length\cite{Hansen2013}
\begin{equation}
k_{\textrm{D}}^2 = \frac{q^2 n_0}{\epsilon k_{\textrm{B}}T}
\label{eqn:kDebye}
\end{equation}
The normal external electric field breaks the symmetry of the layer. However charge neutrality (Eq.~\ref{eqn:neutral}) imposes  $\phi^*(-z) = - \phi^*(z)$ symmetry on the linear system.   Because if the potential is odd under $z \rightarrow -z $ reflection  the electric field $e(z) = - d \phi(z)/ dz)$ is even and therefore  
\begin{equation}
 \int_{-l/2}^{l/2} \rho(z) dz  = e(l/2) - e(-l/2) = 0 
 \label{eqn:lineutr}
\end{equation}
To appreciate the difference with the GC electric double layer model it is instructive to compare to the neutrality condition for the GC system which is usually written as 
\begin{equation}
  \int_0^\infty  \rho(z) dz = - \sigma
  \label{eqn:edlsig}
\end{equation}
where the wall with surface charge $\sigma$ is located at $z=0$.  $\sigma$ is fixed and  the net excess charge density on the electrolyte side of the GC double layer is bound. The film neutrality condition Eq.~\ref{eqn:lineutr} places no such restriction on the excess charge. The charge induced in a half layer
\begin{equation}
  \sigma = \int_0^{l/2}  \rho(z) dz 
  \label{eqn:filmsig}
\end{equation}
can in principle  take arbitrarily large values for  strong enough fields $e_0$ or a wide enough layer.

These symmetry considerations lead us to a solution for the linearized PB Eq.~\ref{eqn:PBlin} of the simple form
\begin{equation}
\phi(z) = A \left(\frac{k_{\textrm{B}}T}{q}\right) \sinh \left( k_{\textrm{D}} z \right)
\label{eqn:phifilm}
\end{equation}
with an corresponding electric field
\begin{equation}
 e(z) = - A k_{\textrm{D}}\left(\frac{k_{\textrm{B}}T}{q}\right) \cosh \left( k_{\textrm{D}} z \right)
\label{eqn:efilm}
\end{equation}
The dimensionless coefficient  $A$  is to be determined by relating it to the applied field $e_0$.  The boundary at $ z =  l/2$  not helpful here because $e(l/2)$  is not simply equal to $ e_0$. Note however that the field at the center is finite. This field must persist even in absence of ions ($n_0 = 0$). What is left in this limit is the pure polarized solvent with uniform field $(\epsilon_0 /\epsilon) e_0  $.  This suggests to impose the limiting condition
\begin{equation}
\lim_{n_0 \rightarrow 0} e (0) =  \left( \frac{\epsilon_0}{ \epsilon} \right) e_0
\end{equation}
which is satisfied for the potential
\begin{equation}
   \phi(z) = -\frac{\epsilon_0 e_0}{\epsilon k_{\textrm{D}}} \sinh \left( k_{\textrm{D}} z \right)
\end{equation}
and therefore 
\begin{equation}
 e(z) = \frac{\epsilon_0 e_0}{\epsilon} \cosh \left( k_{\textrm{D}} z \right)
\label{eqn:efilm0}
\end{equation}
For a very dilute solution we can Taylor  expand $\phi(z)$ in  $k_{\textrm{D}}$ giving
\begin{equation}
   \phi(z) = -\left(\frac{\epsilon_0  e_0}{\epsilon} \right) z -
   \left(\frac {k_{\textrm{D}} \epsilon_0  e_0}{ 2 \epsilon}\right)  z^2 + \dots
\end{equation}
The leading term is indeed the potential in a uniformly polarized linear dielectric continuum, being our  solvent.

\subsection{Interpretation ionic polarization} 
What can we learn about ionic polarization from the planar polarized  film of electrolyte solved in the linear PB approximation in the previous section? The one-dimensional geometry is clearly a restrictive special case because all fields are either longitudinal or constant. In particular the 1-D displacement field $d$ (the $z$ component) must be uniform across the layer and equal to its value $\epsilon_0 e_0$ in the vacuum outside the film.   This means that the Lorentz relation Eq.~\ref{eqn:Lorentzid} can be written as 
\begin{equation}
\epsilon_0 e_0 = \epsilon_0 e(z) + p_i(z) + p_d(z)
\label{eqn:Lorentz1d}
\end{equation} 
This relation must hold everywhere inside the film, which has been made explicit by displaying the $z$ dependence of the fields.  

Using the constitutive relation Eq.~\ref{eqn:pchie} for $p_d$ which remains valid at finite ion concentration Eq.~\ref{eqn:Lorentz1d} can be written as
\begin{equation}
  p_i(z) = - \left( \epsilon e(z) - \epsilon_0 e_0 \right)
  \label{eqn:pi1d}
\end{equation}
Without solvent ($\epsilon = \epsilon_0$) Eq.~\ref{eqn:pi1d} reduces to $p_i(z) = - \epsilon_0 \hat{e}(z)$.
This is the expected behaviour of polarization in a non-uniform dielectric. Comparing ionic polarization to dipolar polarization is justified in this case.  Trivially this also makes sense in the opposite limit of zero ion concentration because $\epsilon e = \epsilon_0 e_0$ for a pure uniform slab of dielectric and Eq.~\ref{eqn:pi1d} sets $p_i(z)$ to zero. At finite concentration and solvent susceptibility  $p_i$ 
evidently makes up the mismatch between $ \epsilon e(z) $ and $\epsilon_0 e_0$  which is at least consistent with the interpretation $p_i$ as a polarization. 

Finally combining Eq.~\ref{eqn:pi1d} with Eq.~\ref{eqn:efilm} we obtain for the ionic polarization  profile
\begin{equation}
p_i(z) = - \epsilon_0 e_0 \left( \cosh \left( k_{\textrm{D}} z \right) - 1 \right)
\label{eqn:pifilm}
\end{equation} 
The ionic polarization is canceled midway the two surfaces  ($p_i(0) = 0$) in contrast to the electric field which is finite due to solvent polarization. Going out to the boundary the ionic polarization increases exponentially with the characteristic length of $1/k_{\textrm{D}}$. Reassuringly the ionic polarization shows  similar behaviour as the excess charge density as it should according to its definition Eq.~\ref{eqn:pion}. In fact inserting Eq.~\ref{eqn:pifilm} and integrating gives for the total charge Eq.~\ref{eqn:filmsig} induced in a half layer  of a width well exceeding the Debye length 
\begin{equation}
 \sigma =  p_i(0) - p_i(l/2) = \epsilon_0 e_0 \cosh \left( k_{\textrm{D}} l/2 \right) 
\end{equation}
which makes the point of the qualitative difference between  a polarized film and GC electric double layer (Eq.~\ref{eqn:edlsig}). This, of course, raises the question about ionic polarization in a GC diffuse layer.  Investigation of diffuse double layers and other geometries (cylinders,wedges) will be deferred to a future more applied publication.  
 
\section{Summary and discussion} \label{sec:disc}
In the conventional picture of the electrostatics of Poisson-Boltzmann theory the divergence of the dielectric displacement field is equal to the density of mobile charge which is treated as external charge.  This is however not the only option. Mobile charge can also be regarded as internal charge which is allowed in Maxwell-Lorentz continuum theory. In this picture the ionic charge density is (minus) the divergence of ionic polarization. With this additional field,  introduced just for this purpose, the dielectric displacement of the electrolyte becomes fully transverse and the ionic solution can be formally regarded as a pure dielectric continuum.  Exploiting this parallel we have applied a  variational method developed for pure dielectric material\cite{Ericksen2007}  to the Poisson-Boltzmann continuum and were able to reproduce the familiar equilibrium equations for electrostatic potential and charge densities from the corresponding dielectric  Euler-Lagrange equations.  

A distinctive feature of the method is the unified treatment of the electrostatic interactions between ions, between  ions and  explicit solvent polarization and the solvent polarization with itself.   This is all taken care of by a single electrostatic field energy density  coupling the sum of ionic and solvent polarization to the dielectric displacement(Eq.~\ref{eqn:Ekel}).  Superposition of ionic and solvent polarization is a reflection of their mutual screening and is explicitly incorporated in the electrostatic interaction Eq.~\ref{eqn:Ekel}.  The field energy is the electrostatic energy of the system in vacuum. The dielectric response of the solvent is accounted for by adding a polarization energy containing the constitutive parameter for solvent polarization (the susceptibility).  Such a separation between electrostatic and constitutive energy terms is characteristic  of continuum electromechanics and considered crucial in this field\cite{Ericksen2007,Ogden2017,Sprik2020}.  This may be an advantage when spatial variation of the solvent dielectric constant is important as has been argued for the modeling of colloidal and macromolecular systems\cite{Cruz2012,Holm2014,Holm2015,Luijten2014}.  Finally, what may be seen as a somewhat eccentric feature, the method introduces a vector potential potential as an auxiliary variational degree of freedom and is therefore inherently a vector field based theory.  Without extensive application in computations involving complex systems  it is therefore not clear how efficient this method will be compared to other schemes\cite{Dunweg2004,Holm2006,Cruz2012,Holm2014,Luijten2014}.  

Regarding formal theory, the change of perspective could be of use in some applications.  One such area is electromechanics. As already mentioned in the introduction, the Ericksen scheme used here was developed for the derivation of stress tensors of dielectric continua. In a recent publication\cite{Sprik2020} the author repeated this derivation employing the systematic Lagrangian electromagnetic formalism of Dorfmann and Ogden\cite{Ogden2017,Ogden2005}. Stress tensors in continuum mechanics are obtained as (non-linear) strain derivatives of an energy functional.  In contrast, the stress tensors in most studies based on density functional theory are integrals of the force density\cite{Evans1979}.  The electric part of the PB stress tensor obtained by this route is essentially the Maxwell stress tensor formed from the Maxwell field\cite{Hansen2013,Evans1980}.  It would be of interest to verify whether the stress tensor formally derived from  the PB functional using the methods of continuum mechanics is indeed equal to the DFT  stress tensor. For example, one of the questions is the effect solvent stress (the Korteweg-Helmholtz stress tensor or a variant thereof\cite{Landau1984,Ericksen2007,Suo2008,Sprik2020}) which is non-trivial even in the simple polarizable continuum model used in PB theory\cite{Bazant2020}.  The reformulation in terms of a functional of  ionic and solvent polarization  should make the PB theory of electrolytes amenable to the established continuum mechanics methods of electroelasticity.   

A further potential application is non-equilibrium thermodynamics.  The point of departure for transport theory and non-equilibrium thermodynamics for electrolytes is the  Poisson-Nernst-Planck (PNP) formalism. This is a huge field with many important practical applications\cite{Bazant2007,Bazant2010,VanRoij2016,Yochelis2018}(to mention just a very small selection). Polarization can be equated to the time integral of  electric current in the quasi electrostatic limit.  This relation was the vital ingredient for defining uniform electronic polarization\cite{Resta1994,Resta2010}  in periodic systems.   Equilibrium polarization in this picture is the left over of transient current. The corresponding definition of global ionic polarization in classical systems under periodic boundary conditions\cite{Sprik2018,Zhang2020b} was used in a recent molecular simulation of  a bulk electrolyte (aqueous NaCl)\cite{Cox2019}. The generalization to local ionic polarization may seem artificial in static systems,  but is more natural in time dependent systems.  The resulting PB functional, while equivalent to the original, seems closer to the structure of PNP theory\cite{Yochelis2018} and may  therefore be more suitable for a time-dependent generalization of the PB functional.  

As a final  observation we point out that the issue of the correct  form of the stress tensor and the transport properties of the PB electrolyte are linked by continuum thermomechanics. The stress tensor enters in  the balance laws for energy and the second law for local entropy production. However the continuum thermomechanics of electroactive systems  remains a controversial issue raising  a number of fundamental questions.   In fact it was here that the renewed focus on Maxwell-Lorentz continuum theory promoted in the Kovetz book  had a most stimulating impact\cite{Kovetz2000,Ericksen2007b,Steigmann2009}. The judgment is still out whether the ``dielectric''  view of PB theory proposed in this paper is more than a theoretical curiosity. However we hope that it could be of use, if not in practical calculations, then at least as a tool for exploring the continuum electro-thermomechanics of the PB electrolyte.     

%

\end{document}